\documentstyle[11pt,psfig]{article}
\begin{document}

\title{Kaon electromagnetic production: constraints set by new 
data\footnote{talk given at the VIII International Conference on 
Hypernuclear \& Strange Particle Physics, Jefferson Lab, Newport 
News, Virginia, U.S.A., October 14-18, 2003.}}

\author{P. Byd\v{z}ovsk\'{y} and M. Sotona\\
{\it Nuclear Physics 
        Institute, \v{R}e\v{z} near Prague, Czech Republic}}
\date{}       

\maketitle

\begin{abstract}
The CLAS data on the photo-production of K$^+$ off the proton are 
utilised to study reaction mechanism of the process in frame of 
the isobaric approach. The missing $D_{13}$ resonance is shown to 
be important for successful description of the data in the whole 
kinematical region. Constructed models provide satisfactory 
predictions for the process. 
\end{abstract}

\section{INTRODUCTION}
Many attempts to describe the electro-magnetic production of K$^+$ 
have been made in the last decades. Most of the models have, however, 
problems with a reasonable description of new experimental data 
collected in JLab and Bonn \cite{preprint}. Among the models which can 
still provide a satisfactory prediction in comparison with new data 
are the latest isobaric ones, Saclay-Lyon \cite{SL96,SLA98}, 
Kaon-MAID \cite{Ben99}, and Janssen et al \cite{Jan}.

With the very latest data on the photo-production of K$^+$, provided  
by the CLAS \cite{CLAS} and SAPHIR \cite{SAPHIR} Collaborations, which 
reveal clearer and more pronounced resonance structure, one can test 
the models more thoroughly and learn more about the reaction mechanism 
of the process, e.g. a relevance of the ``missing'' $D_{13}$ resonance 
can be addressed or manifestations of other exotic objects can be 
looked for.

A good understanding of the electro-magnetic production of K$^+$ off 
the nucleon is also necessary before more complex calculations 
of the hypernucleus production are performed (see contribution by 
T. Motoba). 
In this respect a reliable prescription for the elementary amplitude 
is desirable at very small kaon lab angle ($\theta_K^{lab} < 10$ deg) 
since particularly this kinematical region contributes dominantly to 
production rates of the hypernuclei. 

In this note we limit ourselves to an analysis of the CLAS data set.
Our aim is to find  an isobaric model being in a better agreement with 
the CLAS data than the older modes are. We shall in addition try to 
answer the questions whether the data necessitate an introduction 
of the $D_{13}$ resonance in models and which role does play 
the $N^*(1710)$ resonance, a possible candidate for the 
penta-quark \cite{Wilczek}.

\section{FORMALISM OF ISOBARIC MODELS}
In our analysis we utilise the isobaric approach which uses 
the Feynman diagrammatic technique limited to the exchange of one 
particle or resonance \cite{preprint,SL96}. 
The hadron structure in the strong vertexes is taken into account 
by means of the dipole type form factors \cite{Jan} introduced in 
the gauge-invariant way as proposed by Davidson and Workman \cite{DW01}. 

Besides the Born terms and two kaon resonances, 
K$^*$(891) and K$_1$(1273), we consider exchanges of nucleon: 
$S_{11}(1650)$ (N4), $P_{11}(1710)$ (N6), and $P_{13}(1720)$ (N7), 
and hyperon:
$S_{01}(1670)$ (Y2), $S_{01}(1800)$ (Y3), $P_{01}(1600)$ (Y4), 
$P_{01}(1810)$ (L5), and $P_{11}(1660)$ (S1) resonances.  
Moreover, we consider also the $D_{13}(1895)$ one       
predicted by the quark model and already assumed in the Kaon-MAID 
model \cite{Ben99} but reliably not observed yet.  
The spin 3/2 and 5/2 resonances are treated in the method  
used in \cite{SL96} without off-shell effects \cite{SLA98}.

Parameters of the models, the coupling constants and cut-offs 
of the hadron form factors, were fixed via fitting the differential 
cross section and polarization to the CLAS data set which consists 
of 918 data points for the cross section and 220 points for 
polarization. 
The two main coupling constants were always kept in 
the limits of the 20\% broken SU(3) symmetry:  
$-4.4 \leq g_{KN\Lambda} \leq -3.0$ and 
$0.8 \leq g_{KN\Sigma} \leq 1.3$. Values of the cut-off masses 
were also confined: $0.6 \leq \Lambda \leq 2.0$ GeV. 

\section{DISCUSSION OF RESULTS AND CONCLUSIONS}
The Saclay-Lyon models SL \cite{SL96} and SLA \cite{SLA98}  
describe satisfactorily the old data except at 
$\theta_K^{cm} > 130$ deg and $E_{\gamma}^{lab} > 1.5$ 
GeV \cite{preprint}. The models include nucleon resonances 
with spin up to 3/2 (SLA) and 5/2 (SL) and many hyperon ones. 
Even though the $\chi^2$ improved after re-adjustment 
of the parameters, 398 $\rightarrow$ 5.7 (SL) and 412 
$\rightarrow$ 7.2 (SLA), the models do not predict any resonant 
structure above 1.2 GeV. 
They are able to describe the global behaviour of the data except 
at $E_{\gamma}^{lab}>2$ GeV and small kaon angle where they 
over-predict the cross sections. This is improved when the 
hadron form factors (h.ff.) are included in the models which 
results, e.g., in $\chi^2$ = 4.13 for SLA and damping the cross 
section at $E_{\gamma}^{lab}>1.7$ GeV and small kaon angle.
Absolute values of some hyperon coupling constants of the model 
with the h.ff., however, tend to be large ($\approx9$) and their 
uncertainty due to the fitting procedure is also big ($\approx$ 
100-130\%). Predictions of the polarization are not good with 
these re-fitted Saclay-Lyon models. 

%
%
\begin{figure}[htb]
\centerline{\psfig{figure=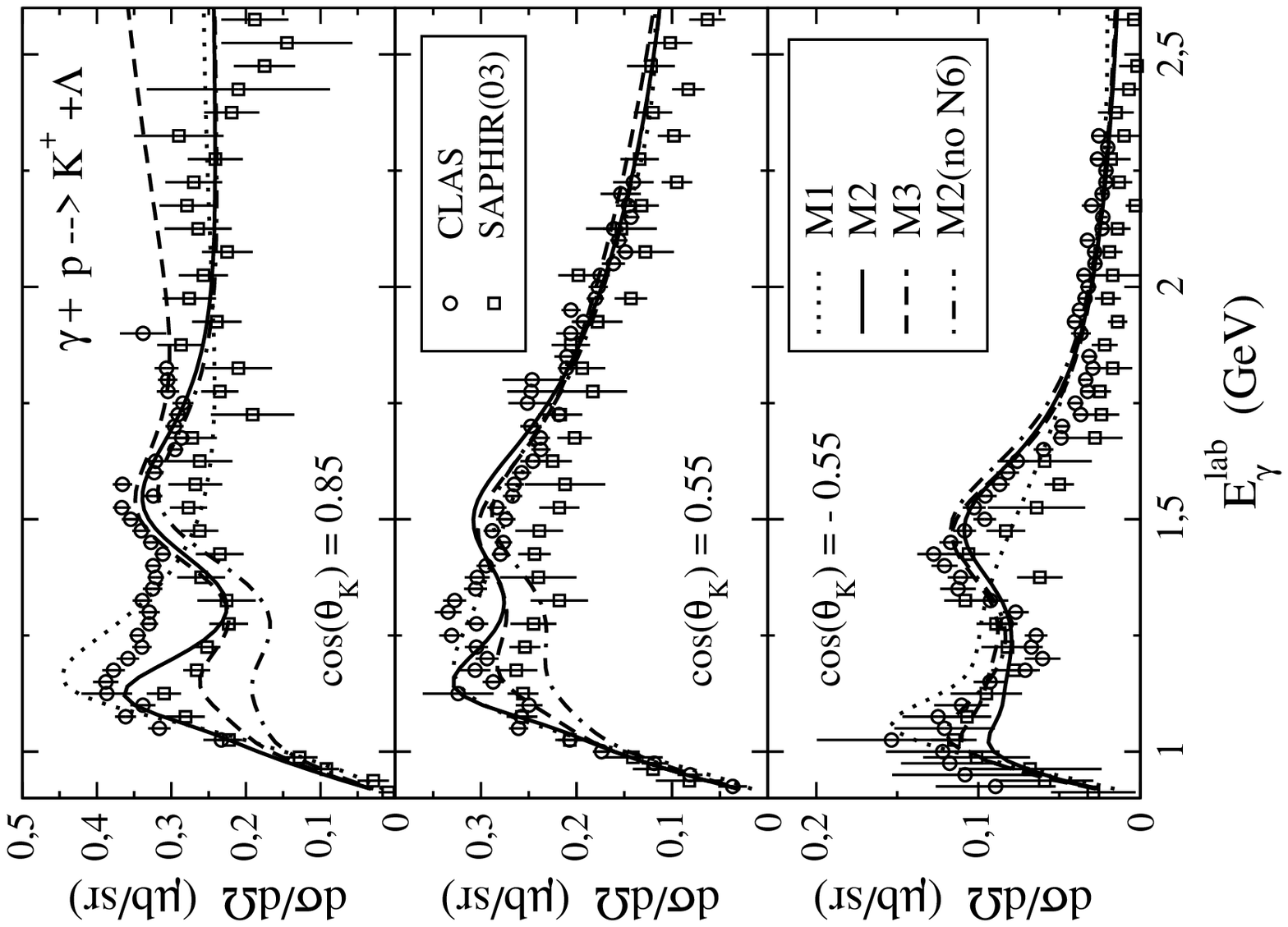,angle=270,width=76mm}
\psfig{figure=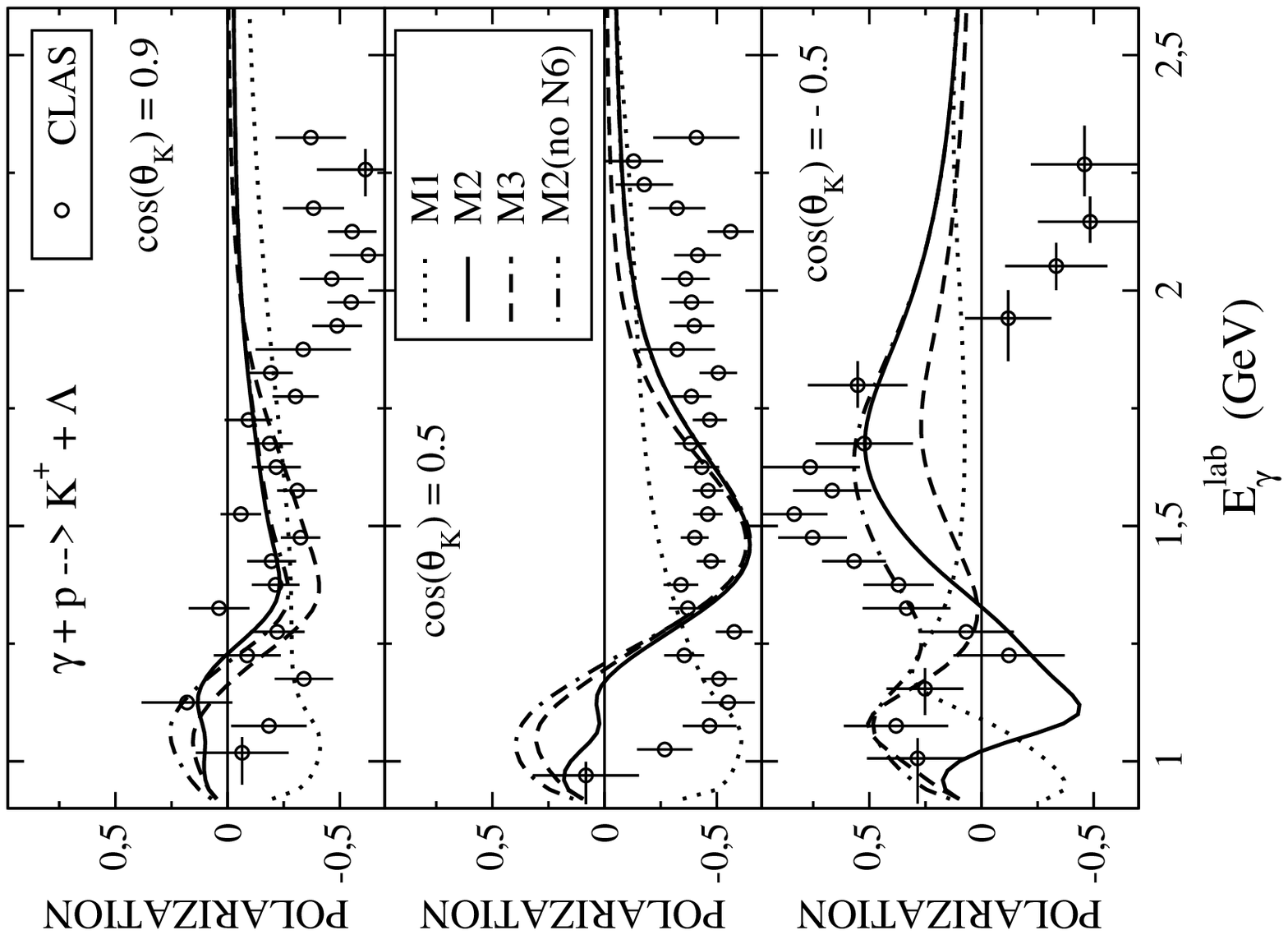,angle=270,width=76mm}}
\vspace*{-5mm}
\caption{Predictions of the models M1, M2, and M3 are compared 
with data \cite{CLAS,SAPHIR}.\vspace{-3mm}}
\label{fig1}
\end{figure}
The Kaon-MAID model \cite{Ben99} includes in addition to 
the Born, K$^*$, and K$_1$ terms only the nucleon resonances 
N4, N6, N7, and $D_{13}$. We have chosen three modifications 
of the model which differ in the resonance content:
M1 - without $D_{13}$; M2 - all the above resonances; 
M3 - without N6. In these models the h.ff. of Davidson-Workman 
type are assumed with independent cut-offs for the Born and resonant 
terms. The $\chi^2$ for the models are: 2.85 (M1), 2.70 (M2), 
and 3.40 (M3). The coupling constants acquire reasonable values. 
Results for the differential cross sections and polarizations are 
presented in Fig.1 in comparison with the CLAS and SAPHIR data.    
The model M1 predicts no resonant structure in the cross section  
above 1.2 GeV at forward and backward kaon angles but it does predict 
a second maximum at 1.3 GeV and around $\theta_K^{cm}$ = 105 deg. 
Good results are provided by the model M2 except for forward angles   
where it predicts lower cross sections around 1.3 GeV than the CLAS 
data suggest. At the first maximum ($E_{\gamma}^{lab} \approx 1.15$ 
GeV) the model M3 gives smaller cross sections at forward angles 
($\cos(\theta_K)$ = 0.85) whereas it gives larger values at backward 
angles ($\cos(\theta_K)$ = -0.55) than the M2 model. 
The model M2 without the N6 resonance displays a similar but still 
more pronounced pattern (dash-dotted line). The effects were not 
seen, however, when we omit N6 and add two hyperon resonances 
(Y2, L5) or (Y2, Y4). Then the first peak can be modelled 
satisfactorily too. On the contrary, the N6 resonance alone 
is not able to mimic the peak sufficiently even though various 
hyperon resonances are added (Y2, Y4, L5, S1). An absence of N6 in 
the model M2 results in the opposite sign for polarization around 
1.15 GeV and large angle ($\cos(\theta_K)$ = -0.5). 
The model M1 predicts different behaviour of the polarizations 
in comparison with the models M2 and M3. All models do not predict 
proper values for the polarizations above 1.7 GeV and small kaon 
angle (Fig. 1 right panel). 

%
%
\begin{figure}[htb]
\centerline{\psfig{figure=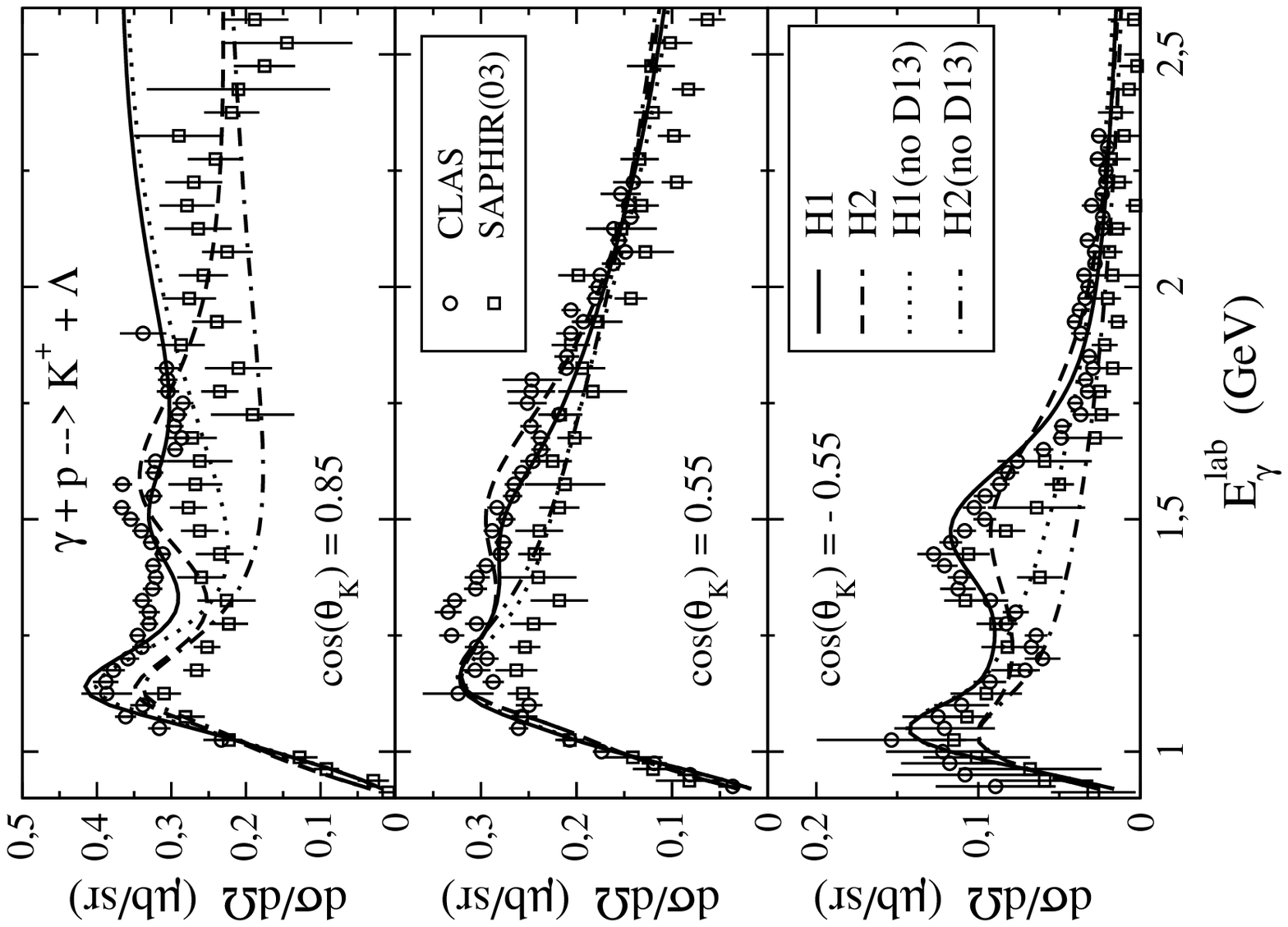,angle=270,width=76mm}
\psfig{figure=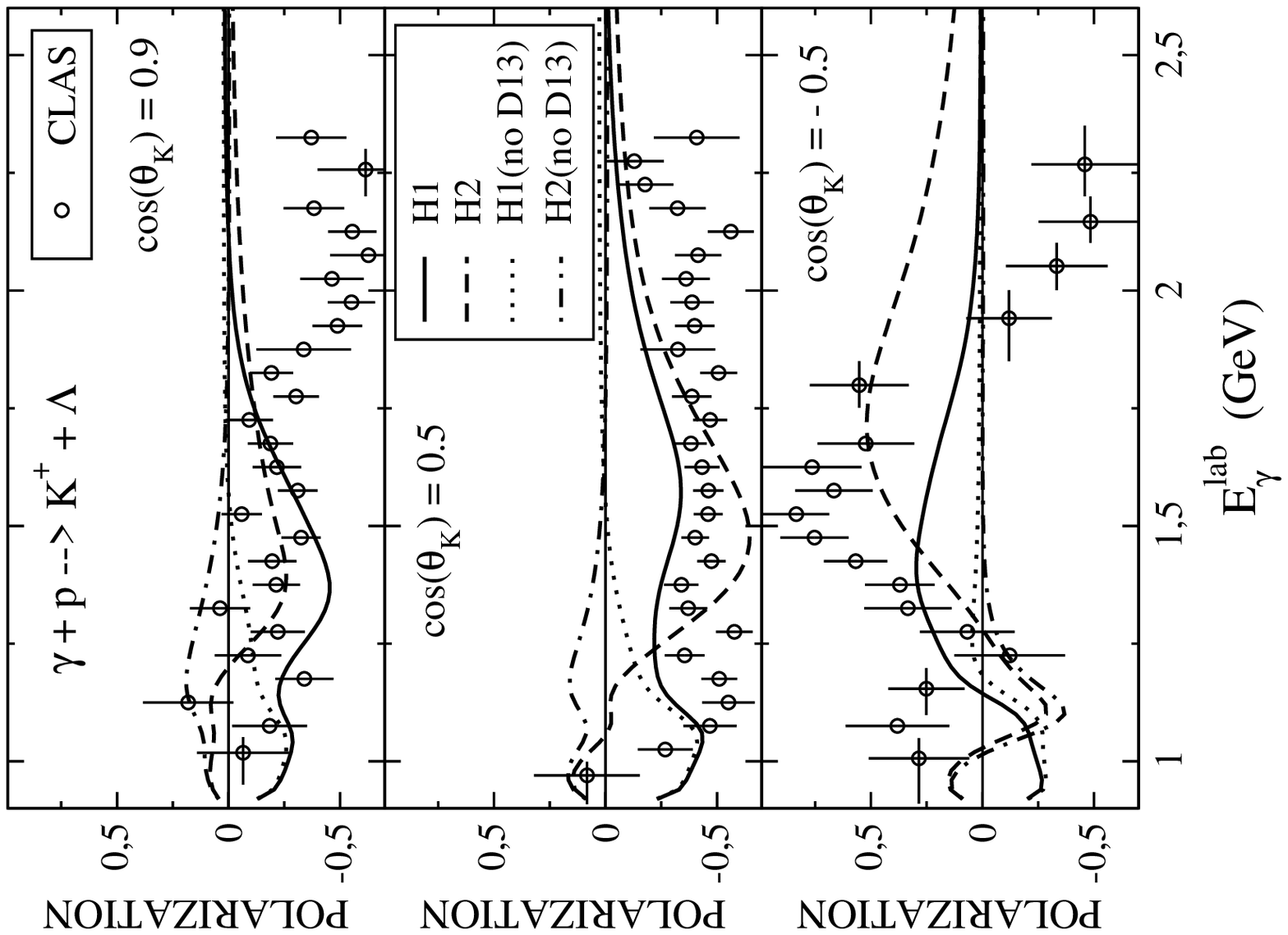,angle=270,width=76mm}}
\vspace*{-5mm}
\caption{Predictions of the models H1 and H2 are compared with data 
\cite{CLAS,SAPHIR}.\vspace{-3mm}}
\label{fig3}
\end{figure}
In Figure 2 the best models of those we tried with hyperon resonances 
are shown. In addition to the resonances contained in M2 these models 
include: Y2 and L5 (model H1); Y2 and Y3 (model H2). 
Here a common cut-off for all terms is assumed in the h.ff. 
Those versions of the model in which the Y2 resonance is replaced 
by Y3, Y4 or S1 provide very similar results. 
The $\chi^2$ for the models H1 and H2 are 2.55 and 2.74, respectively. 
Predictions of these models for the cross sections differ around the 
first and second maximum (Fig. 2). This illustrates that the resonance 
pattern is sensitive to a choice of the hyperon resonances too (Y3 and 
L5 have almost equal mass but opposite parity). 
Both models fail to describe the polarization above 1.8 GeV and small 
kaon angle like the model M2 (see top part of the right panels in 
Figs. 1 and 2). 

Results of the models without $D_{13}$ (dotted and dash-dotted lines 
in Fig. 2) show a significance of the $D_{13}$ contribution to the 
cross section in energy region 1.3 - 1.8 GeV~for all angles.
This observation and that for the model M1 suggest importance 
of $D_{13}$ for a proper description of the cross section at the 
second maximum. A particular shape of the structure can be, however, 
influenced by a presence of hypernuclear resonances. 
In Figure 2 it is shown that $D_{13}$ also dominates 
the polarizations for $E_{\gamma}^{lab}> 1.5$ GeV in the models. 

The hyperon coupling constants for the H1 and H2 models are 
large with big error bars like they are for the re-fitted Saclay-Lyon 
models. Small cut-off parameters, 0.73 GeV (H1) and 0.88 GeV (H2), 
mean that soft h.ff. are present in these models. 

We conclude that in the isobaric models, in which the off-shell 
effects connected with the spin 3/2 resonances are not assumed, 
the $D_{13}$ resonance has to be included to describe the resonant 
structure of the cross sections correctly. This resonance can mimic 
the second maximum at all angles in interference with the background 
which need not include the hyperon resonances. A presence of the 
$N^*$(1710)(N6) resonance in the models makes description of the 
cross section easier at the first maximum but it is not necessary 
to include this resonance in the models.

\section*{ACKNOWLEDGEMENTS}\vspace{-2mm}
We are much obliged to R. Schumacher and K.-H. Glander 
for providing us with the CLAS and SAPHIR data, respectively. 
This work was supported by grant GACR202/02/0930.


\end{document}